\documentclass[prb,twocolumn,floatfix,amsmath,amssymb]{revtex4}
\usepackage[dvips]{graphicx}
\usepackage{latexsym}
\usepackage{graphicx}
\usepackage{times}
\usepackage{amsmath}
\usepackage{dcolumn}
\usepackage{latexsym,amsmath,amssymb,bm,euscript}
\usepackage{bm}
\usepackage{amsmath}
\usepackage{euscript}
\usepackage[dvips]{graphicx}
\usepackage{hyperref}
 \hypersetup{%
    breaklinks = {true},
    citecolor = {blue},
    colorlinks = {true},
    linkcolor = {red},
    pdfcreator = {\LaTeX\ and \flqq hyperref\frqq},
    pdffitwindow = {true},
    pdfmenubar = {true},
    pdfpagelayout = {SinglePage},
    pdfstartview = {Fit},
    pdftoolbar = {true},
    plainpages = {false},
  }
\begin{document}
\bibliographystyle{apsrev}

\title{Weak Antiferromagnetic Order in  Anisotropic Quantum  Pyrochlores}

\author{Valeri N. Kotov} 

\affiliation{Department of Physics, Boston University, 590 Commonwealth Avenue, 
Boston, Massachusetts  02215}

%\date{\today}
\begin{abstract}
 We study the ground state properties of an anisotropic,
 quasi-2D version of the quantum (S=1/2) pyrochlore antiferromagnet.
 In the presence of Dzyaloshinsky-Moriya interactions, in addition to the 
 Heisenberg exchanges, it is shown that two types of ordered magnetic states are 
 generally possible:  non coplanar ``chiral," 
 and coplanar antiferromagnetic order. The magnetic moments in 
 all cases are determined by the Dzyaloshinsky-Moriya interactions and in
 this sense the antiferromagnetic order is ``weak."
\end{abstract}
%\pacs{
%To be determined later
%75.10.Jm, 75.30.Ds 
%}
\maketitle

\section{Introduction and Description of the Model}

The pyrochlore antiferromagnet is strongly frustrated and the structure
 of its ground state represents a challenging problem in the field of magnetism.
It has been argued that, for purely Heisenberg interactions, 
both the classical and quantum versions  of the model
 are magnetically disordered.\cite{MCCC,HBB,T}
It also appears that in the extreme quantum limit (S=1/2),
 lattice-symmetry breaking and spontaneous dimerization take
 place, although the ground state still exhibits macroscopic
degeneracy.\cite{T,BAA,MSG} Thus it  is natural  to expect that
interactions beyond the Heisenberg exchange, as well  as inclusion
 of various magnetic anisotropies, can  impose their own order,
 such as orbital, dimer, or  magnetic, or combinations of these.
  Effects due to orbital degeneracy, \cite{MJP}
 long-range dipolar interactions, \cite{PC} and spin-lattice interactions
  \cite{TMS1} have been studied. Ising anisotropies can 
 lead to the formation of ``spin ice", \cite{BG} which bears resemblance
 to the problem of proton disorder in ice.\cite{APF}
  Various planar \cite{FMSL,TYM} and other anisotropic \cite{SFB} versions
 are also of strong theoretical  interest, although they do not necessarily
 reflect the physics of the full 3D pyrochlore structure. 
 Finally, under certain conditions, even more exotic ground states have
 been proposed, such as U(1) spin liquids, \cite{HFB}  or non-magnetic
 chiral states.\cite{KH}
 
 In this work we will discuss  certain aspects of the mechanism for magnetic 
 order formation due to the presence of  Dzyaloshinsky-Moriya (DM) 
 interactions.\cite{DM}
  For the case of classical spins,  this mechanism was studied recently \cite{MCSL}
 (see also \cite{CEL}). The  quantum case (S=1/2) was discussed \cite{KEZM}
 within a technical scheme, similar to the one used in Ref.~\onlinecite{T} for
 the pure Heisenberg case. A puzzling difference between the classical
 and the extreme quantum case is that in the classical case 
 non planar, chiral-like (see below)   as well as  coplanar spin order is 
 possible,\cite{MCSL} while in the quantum case only the chiral order
 was predicted.\cite{KEZM} Naturally, the way spin order emerges is also
 quite different in the classical and quantum cases,   and for S=1/2 the
 induced  antiferromagnetic order is ``weak," in a sense that the ordered
 moment is proportional to the DM interaction itself.
 We point out that the role of DM interactions has been extensively studied
 only in the context of non-frustrated lattices, such as the square lattice,
 where it typically  leads to  weak ferromagnetic moments (present, for example,
 in the copper oxide compounds.) In frustrated lattices, however, the DM interaction
 effects can be much more profound, and are expected to 
 lead to complex types of order.\cite{CB}

%--------------------
%Fig1
%--------------------
\begin{figure}[ht]
  \centering
  \includegraphics*[width=0.45\textwidth]{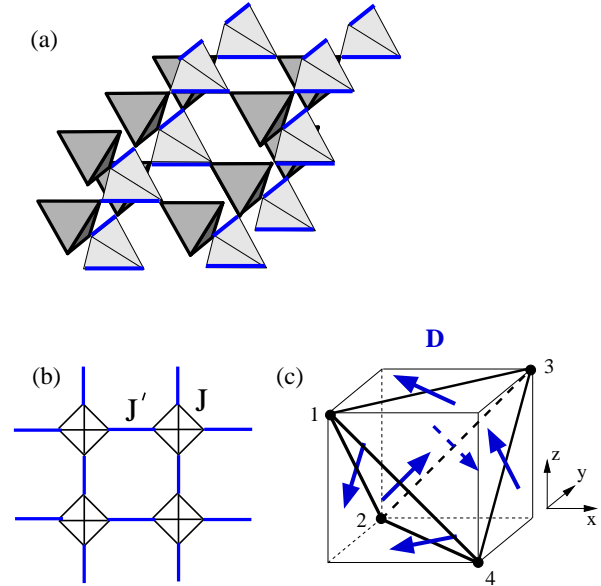}
  \caption{
(Color online) (a) Pyrochlore lattice (not all tetrahedra shown, for clarity.)
 (b) 2D Pyrochlore ``layer,"  showing
 the bonds in a $(001)$ slice of the pyrochlore lattice. (c) DM interactions
on a tetrahedron. The DM vectors are represented by blue arrows.
  }
  \label{Fig1}
\end{figure}

 We consider the anisotropic version of the 3D pyrochlore lattice (Fig.~\ref{Fig1}(a)), where
 it can be viewed as weakly-coupled ``layers," defined  in Fig.~\ref{Fig1}(b).
 Our goal is to investigate the possible types of magnetic order arising in this situation.
 It is technically very advantageous to consider the anisotropic limit, namely
 $J_\perp \ll J' \ll J$,  where  $J_\perp$ is the Heisenberg exchange between the layers
(thin black lines in Fig.~\ref{Fig1}(a)), and $J'$ is the inter-tetrahedral exchange
 (solid blue lines in  Fig.~\ref{Fig1}(a,b)).  $J$ is the exchange on the ``strong" (shaded) 
 tetrahedra, shown as plaquettes in Fig.~\ref{Fig1}(b). This strong-coupling
 approach is similar to the one used earlier,\cite{T,BAA,KEZM} except that now we
 consider an anisotropic version of it. The quasi 2D version  allows us also
 to monitor the low-energy excitation spectrum, and determine the conditions
 under which different types of DM-induced order can arise. In order to implement this
 program we need to know the exact excitation spectrum on a single tetrahedron,
 which we calculate below, and then consider the lattice version of coupled tetrahedra.
 The lattice shown on Fig.~\ref{Fig1}(b) is a frustrated one,  and we will present
 arguments why the strong-coupling expansion,  governed by the parameter $J' < J$,
 should work well in determining the ground state structure.
 In order to avoid cumbersome formulas we present below  results for the 
 strict limit $J_\perp =0$,  while  we have checked that the types of order we find
 are pretty generic (as long as the system is away from the strictly  isotropic 3D case,
 which requires different considerations.) Notice also that the lattice of Fig.~\ref{Fig1}(b)
 is not the same as the 2D projection of the 3D pyrochlore, i.e.
 the checkerboard lattice,\cite{FMSL} which has  valence-bond solid order
(expected, quite generally, to compete with DM-induced magnetic order.)
 
The Hamiltonian reads
\begin{equation}
{\cal H} = \sum_{{\bf i,j}}J_{{\bf i,j}} {\bf S}_{\bf i}.{\bf S}_{\bf j}
 + \sum_{{\bf i,j}}{\bf D}_{{\bf i,j}} .({\bf S}_{\bf i} \times {\bf S}_{\bf j}),
\label{ham}
\end{equation}
\noindent
where the couplings $J_{{\bf i,j}}$ are assumed to be antiferromagnetic
($J_{{\bf i,j}}>0$) and are distributed  as already discussed. All spins are S=1/2. 
 The DM  vectors ${\bf D}_{{\bf i,j}}$ on a tetrahedron are shown in Fig.~\ref{Fig1}(c),\cite{MCSL}
 and have equal magnitude  $|{\bf D}_{{\bf i,j}}|=|D|$.
 Two patterns are possible: the one shown on the figure, and  a pattern with
 all DM vectors reversed, ${\bf D}_{{\bf i,j}} \rightarrow -{\bf D}_{{\bf i,j}}$.
  There is no reason to expect that the induced order  in the two cases will be the same; in fact
 two different types of order were found in the classical version.\cite{MCSL}

In the rest of the paper, we  first calculate the exact spectrum on a single
 tetrahedron (Section II), which is then used  to analyze the spectrum on the 
lattice and determine the possible types of order in Section III. Section IV contains
 our conclusions.

\section{Quantum spins on  a  tetrahedron with DM Interactions}

Without DM interactions, the ground state on a tetrahedron
 of spins S=1/2 is a twofold degenerate singlet. We denote the two states
by $|s_{1} \rangle,|s_{2} \rangle$. Their explicit definition is 
given in Appendix A. The spectrum above the ground state consists
 of  3 degenerate triplets
 ($|p_{\alpha} \rangle,|q_{\alpha} \rangle,|t_{\alpha} \rangle$, \ $\alpha=x,y,z$),
see Appendix A,  as well as S=2 states which are irrelevant for our purposes.
 However, the DM interactions break spin-rotational symmetry, and thus lead to mixing 
 of singlet and triplet
 states. The ground state is still degenerate, and the two new  ground states are:
\begin{eqnarray}
\label{gs}
 |\Psi \rangle  & =  &  
 \alpha |s_{2} \rangle \nonumber\\
&& + \frac{i\beta}{\sqrt{3}} \left \{ |p_{x} \rangle +  |p_{y} \rangle
+  |q_{x} \rangle -  |q_{y} \rangle + 2\sqrt{2}|t_{z} \rangle  \right \}, \nonumber \\ 
 |\Phi \rangle &  = &   \alpha |s_{1} \rangle +i\beta \left(|p_{x} \rangle -  |p_{y} \rangle
 + |q_{x} \rangle +  |q_{y} \rangle \right).
\end{eqnarray}

From now on we measure all energies in units of $J$, i.e. we set
 $J=1$, and consequently $J'/J \rightarrow J', \ D/J \rightarrow D$, etc.
 The ground state energy, corresponding to the states \eqref{gs} is
\begin{equation}
E_{0}= -1 + \frac{\sqrt{2}}{4}D - \frac{1}{4} \left[ 4 + 4 \sqrt{2} D
+ 26 D^{2}\right]^{1/2} .
\end{equation}
For small $D \ll 1 $, 
\begin{equation}
E_{0} \approx -\frac{3}{2} - \frac{3}{2} \ D^{2} .
\end{equation}
The coefficients $\alpha, \beta$ in the wave-functions \eqref{gs}
are given explicitly by the formulas
\begin{equation}
\label{cof}
\alpha = \frac{A}{\sqrt{A^{2}+4}}, \  \  \beta = \frac{\alpha}{A},
\  \ A \equiv - \frac{\sqrt{6}D}{3/2  + E_{0}}.
\end{equation}
It is also useful to have the expansion for small $D$,
\begin{equation}
\label{cof1}
A \approx \frac{2\sqrt{6}}{3} \frac{1}{D}, \ \ \alpha \approx \mbox{sign}(D), \ \
\beta \approx \frac{3}{2\sqrt{6}} |D|.
\end{equation}

In our convention, the positive sign of $D$, $D>0$ in all the above equations
 corresponds to the pattern shown in Fig.~\ref{Fig1}(c)
(called ``indirect" in Ref.~\onlinecite{MCSL}), while $D<0$ (``direct" case from
  Ref.~\onlinecite{MCSL}) is the situation when  all arrows in Fig.~\ref{Fig1}(c) are 
 reversed. $|D|$ is the magnitude of the ${\bf D}_{{\bf i,j}}$ vectors.

%------------------
%Fig2
%------------------
\begin{figure}[tb]
  \centering
  \includegraphics*[width=0.38\textwidth]{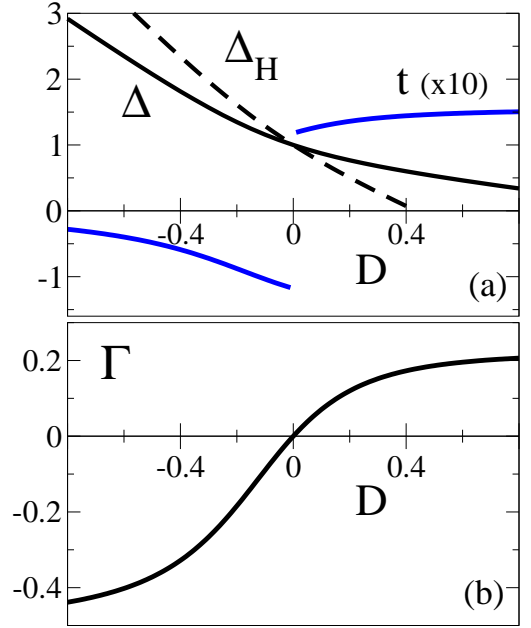}
  \caption{(Color online) (a) The gaps $\Delta = E_b -E_0$ (solid line), 
 $\Delta_{H}= E_{b}(H) -E_0$ (dashed line), and the  
``hopping" parameter $t$ (Eq.~\eqref{t}), solid blue line,  as a function of the DM interaction strength.
(b) The parameter $\Gamma$ (Eq.~\eqref{gama}) which determines the magnitude
 of the ordered moments. $D$ is in units of $J=1$.}
  \label{Fig2}
\end{figure}

Next, the first excited state (which we call $|b \rangle$) is
\begin{equation}
\label{ex1}
 |b \rangle = \frac{1}{\sqrt{6}} \left \{ |p_{x} \rangle +  |p_{y} \rangle
+  |q_{x} \rangle -  |q_{y} \rangle - \sqrt{2}|t_{z} \rangle  \right \},
\end{equation}
with the exact energy
\begin{equation}
E_{b}=  - \frac{1}{2} -  \sqrt{2} D.
\end{equation}
The next excited  state is the ``triplet" $|\bf{P}\rangle$ with components:
\begin{eqnarray}
\label{ex2}
 |P_{z} \rangle  & =  &   \frac{1}{2} \left \{-|p_{x} \rangle +  |p_{y} \rangle
+ |q_{x} \rangle +  |q_{y} \rangle \right \},  \\
 |P_{x,y} \rangle  & =  &   \frac{1}{2} \left \{|p_{z} \rangle \pm |q_{z} \rangle +
 \sqrt{2}|t_{x,y} \rangle \right \}. 
\end{eqnarray}
The  corresponding energy is
\begin{equation}
E_{\bf{P}}=  - \frac{1}{2} -  \frac{\sqrt{2}}{2} D .
\end{equation}

For $D=0$ all excited states are degenerate and separated by an
 energy gap $J=1$ from the ground state. However for finite $D$, the $b$-state  
is the lowest (for $D>0$), and the variation of its gap $\Delta  =  E_{b} - E_{0}$
as a function of $D$ is shown in  Fig.~\ref{Fig2}(a). In fact the gap vanishes
 for the (unphysically) large value of  $D = \sqrt{2}$; on the other hand
 the gap increases for $D<0$. We proceed to explore if this difference can
 affect the type of magnetic order on the lattice (i.e. we study the conditions
 under which  $b$ can condense.)

\section{Low-Energy effective lattice theory: chiral versus coplanar spin order}

\subsection{Low-energy dynamics of interacting spins}

The  ground state structure from \eqref{gs} implies that the spin operators 
 have non-zero matrix elements of the type $\langle \Psi |{\bf S}_{i} | \Phi \rangle$.
This is due to the presence of triplet states within the ground state subspace.
 Using the explicit  form for the various triplet-triplet transition matrix elements,
 summarized in \eqref{mel}, we obtain
\begin{equation}
\label{chis}
 {\bf S}_{i} = \Gamma  \  {\bf{\Lambda}}_{i}  \  T_{{\bf i}}^{y} \ ,
\end{equation}
where the vectors ${\bf{\Lambda}}_{i}$ are defined as
\begin{eqnarray}
&{\bf{\Lambda}}_{1}  =  (-1,-1,1), &\ \ \   {\bf{\Lambda}}_{2} = (-1,1,-1),\\ \nonumber
& {\bf{\Lambda}}_{3} =  (1,1,1),&\ \ \  {\bf{\Lambda}}_{4} = (1,-1,-1) \ .
\end{eqnarray}
The site label $i=1,2,3,4$ in ${\bf S}_{i}$, ${\bf{\Lambda}}_{i}$, refers to the numbering of spins  
on a tetrahedron as displayed in Fig.~\ref{Fig1}(c).
 The operator $T_{{\bf i}}^{y}$ is defined
 as (its bold index ${\bf i}$ refers to the whole tetrahedron)
\begin{equation}
 T_{{\bf i}}^{y} = \frac{i}{2} \left ( \Psi^\dagger  \Phi
-  \Phi^\dagger \Psi  \right ) \, ,
\end{equation}
where $\Psi^\dagger, \Phi^\dagger$ are operators that create the two ground states. 
Obviously  $T^{y}$ is the $y$ (``magnetic") component of the 
pseudospin operator ${\bf T}=1/2$, defined
 in such a way that  $T^{z}=\pm 1/2$ label the two ground states
 \eqref{gs}, i.e.   $T^{z} = \frac{1}{2} ( \Phi^\dagger  \Phi
-  \Psi^\dagger \Psi )$. The coefficient $\Gamma$ in  \eqref{chis} depends
 on $D$ through the wave-function coefficients,
\begin{equation}
\label{gama}
\Gamma = \frac{4}{3} \ \alpha \beta - \frac{2}{\sqrt{3}} \ \beta^{2} \approx \frac{2}{\sqrt{6}} \ D,
 \ \ (D \ll 1).
\end{equation}
We have also given the small $D$ expansion which follows from \eqref{cof1}. The plot
 of the exact function $\Gamma = \Gamma (D)$ is shown in Fig.~\ref{Fig2}(b); notice
 that it is not $D\rightarrow -D$ symmetric.

%--------------------------
%Fig3
%--------------------------
\begin{figure}[tb]
  \centering
  \includegraphics*[width=0.38\textwidth]{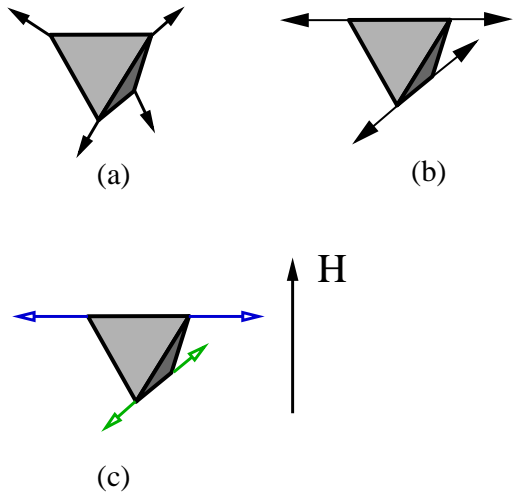}
  \caption{(Color online)  Types of magnetic order: (a) Chiral, (b) Coplanar, and
 (c) Coplanar in an external magnetic field (blue arrows are longer than green ones.) }
  \label{Fig3}
\end{figure}

 In the lattice geometry of Fig.~\ref{Fig1}(b), the effective interaction
 between the  pseudospins is determined by the spin-spin interactions, and
 appears already at order $J'$. From  \eqref{chis} we easily obtain
\begin{equation}
\hat{{\cal H}}_{\mbox{eff}}^{(DM)}=  - J' \Gamma^{2}
\sum_{\langle{\bf i}{\bf j}\rangle}  T_{{\bf i}}^{y}T_{{\bf j}}^{y},
\label{phamDM}
\end{equation}
implying that ferromagnetic ordering takes place (since $J'>0$), i.e. on every
 tetrahedron  $\langle  T_{{\bf i}}^{y} \rangle = 1/2$ 
(or $\langle  T_{{\bf i}}^{y} \rangle = -1/2$). The Ising symmetry 
$T_{{\bf i}}^{y} \rightarrow -  T_{{\bf i}}^{y} $ is spontaneously broken.
 Physically, the state $\langle  T_{{\bf i}}^{y} \rangle = 1/2$ 
 corresponds to a unique ground state formed as the linear
 combination: $ \left (|\Phi \rangle + i |\Psi \rangle \right)/\sqrt{2}$.
 There is magnetic order in this ground state, as follows from
 \eqref{chis}: $\langle {\bf S}_{i} \rangle = (1/2)  \Gamma  {\bf{\Lambda}}_{i}$;
it is  shown in Fig.~\ref{Fig3}(a).
 This is a magnetic ``chiral" state,  characterized by  a non-zero
 scalar chirality $\langle {\bf S}_i\cdot({\bf S}_j \times {\bf S}_k) \rangle \neq 0$.
 The magnitude of the magnetic moment is $|\langle {\bf S}_{i} \rangle| = (\sqrt{3}/2)| \Gamma|$,
 and the order is ``weak" in a sense that $\Gamma$ is determined by the value of the  DM vectors.
  It is interesting to note that chiral order has been discussed  in the context
 of  spin liquid physics, \cite{WWZ,KH}  where one can presumably have a
 state with broken  time-reversal symmetry and yet not magnetic order.
 In our case however  the presence of DM interactions (which break spin-rotation
 symmetry) makes any time-reversal broken state magnetic.
 
 Magnetic chiral spin states were also discussed in the context of the full 3D pyrochlore
 lattice and in other (quasi 2D) non-frustrated situations. \cite{KEZM,KZEM}
 Such states generally compete with dimer order, which technically manifests
 itself in the presence of $T_{{\bf i}}^{x}, T_{{\bf i}}^{z}$ operators in the effective
 Hamiltonian. Averages of such operators in the ground state lead to dimer order,
  and in turn diminish
 the  magnetic, $T_{{\bf i}}^{y}$ component. \cite{T,KEZM,KZEM} 
  The  specifics of this competition depend on the lattice.
 In our case (Fig.~\ref{Fig1}(b)) the issue of spontaneous dimerization
 has not been studied, to the best of our knowledge. However, it is clear
 that couplings involving $T_{{\bf i}}^{x}, T_{{\bf i}}^{z}$ 
appear only in order $(J')^{4}$, and higher.
Thus we will assume  they can be neglected in the limit
 $J' \ll 1$.  Given the coefficient in \eqref{phamDM}, a more precise
 criterion for the magnetic order to be dominant over potential
 spontaneous dimerization is $J' \Gamma^{2} > (J')^{4}$,
 which we implicitly assume to be satisfied.
 Thus  \eqref{phamDM} determines the ground state structure
 within  the degenerate subspace, and the degeneracy is lifted as 
 explained previously, by locking $\langle  T_{{\bf i}}^{y} \rangle = 1/2$,
 which will be assumed from no on. (The choice $\langle  T_{{\bf i}}^{y} \rangle = -1/2$
 leads to time-reversal of all magnetic states.)

Now we write the effective Hamiltonian for the lowest excited state $b$,
 Eq.~\eqref{ex1}. It is convenient to  express the spin operators in the
 ground state $\langle  T_{{\bf i}}^{y} \rangle = 1/2$ via the $b, b^{\dagger}$ operators,
 similarly to the way it is summarized for the case of zero chirality ($D=0$) 
in \eqref{mel}. Performing the necessary calculations, we obtain
\begin{eqnarray}
\label{ops1}
S_{1,3}^x  & = & \pm\frac{t}{\sqrt{2}}\left(1-i \sqrt{3}\right)b^{\dagger} + {\mbox{h.c.}},
 \ \ S_{2,4}^x = S_{1,3}^x,
  \nonumber \\
S_{1,3}^y & = &  \pm\frac{t}{\sqrt{2}}\left(1+i \sqrt{3}\right)b^{\dagger} + {\mbox{h.c.}},
\ \  S_{2,4}^y = - S_{1,3}^y,
   \nonumber \\
S_{1,3}^z & = & \sqrt{2} t b^{\dagger} + {\mbox{h.c.}}, 
\ \  S_{2,4}^z = - S_{1,3}^z.
\end{eqnarray}
\noindent
In \eqref{ops1} we use notation such that the lower left index of
 $S_{i,j}^{\alpha}$ (which stands for either $S_{i}^{\alpha}$ or $S_{j}^{\alpha}$) corresponds
 to the upper sign on the right hand side, and the lower right index---to the lower
 sign (i.e. $S_{1}^x = + \frac{t}{\sqrt{2}}\left(1-i \sqrt{3}\right)b^{\dagger} +{\mbox{h.c.}}$,
 and $S_{3}^x = - \frac{t}{\sqrt{2}}\left(1-i \sqrt{3}\right)b^{\dagger} +{\mbox{h.c.}}$, etc.)
The formulas \eqref{ops1} refer to a single tetrahedron which we label
 (as before) with a bold index ${\bf i}$ (i.e. $b$  will carry this index, $b \rightarrow b_{\bf i}$.)
 The site index $i=1,2,3,4$ of the spins $S_{i}^{\alpha}$ on a tetrahedron
 again follows the convention shown in Fig.~\ref{Fig1}(c).
We also emphasize that equations \eqref{chis},\eqref{ops1}
 simply mean that  the spin operators  have the same matrix elements as
 the right hand sides of those equations,  i.e. the expressions 
  should be added up to obtain the total spin operators.

The coefficient $t$ is defined as 
\begin{equation}
\label{t}
t \equiv  \frac{\alpha}{6\sqrt{2}} +  \frac{\beta}{2\sqrt{6}}  \approx
 \frac{ \mbox{sign}(D)}{6\sqrt{2}} + \frac{|D|}{8}, \ (D\ll1),
\end{equation}
and  the small $D$ limit expansion is also given.
 Figure \ref{Fig2}(a) shows a plot of $t$.

The effective Hamiltonian, describing  the low-energy dynamics of $b$ can
 now be readily obtained:
\begin{eqnarray}
\label{teff}
H_b &=& \Delta  \sum_{{\bf i}} b_{{\bf i}}^{\dagger} b_{{\bf i}}^{}
 +  \sum_{\langle{\bf  i j } \rangle} \left( t_{1} b_{\bf i}^{\dagger}  b_{\bf j}^{} + 
t_{2} b_{\bf i}^{\dagger} b_{\bf j}^{\dagger} + {\mbox{h.c.}} \right) \nonumber \\
 & & +  t_{3} \sum_{\bf i} \left(  b_{\bf i}^{\dagger} + b_{\bf i}^{} \right).
\end{eqnarray}
The parameters appearing in $H_b$ are
\begin{eqnarray}
 \Delta & = & E_{b} - E_{0}, \nonumber \\
 t_{1}  & = & -2 t^{2}(J' + \sqrt{2} D'), \nonumber \\
 t_{2}  & = & 4 t^{2}(J' - \sqrt{2} D'/2), \nonumber \\
 t_{3}  & = & 4\left(\frac{\Gamma}{2}\right) t (2 \sqrt{2} J' +  D').
\end{eqnarray}
Here we have also included the DM interactions $D'$ which appear
 on inter-tetrahedral bonds (and we set $D'=D$ from now on); their distribution
 is explained in Ref.~\onlinecite{MCSL}. The presence of $D'$ is not qualitatively 
(or even quantitatively) important for our following discussion.
 The quantity $\Delta$ is the on-site gap already discussed previously,
 while $t_{1},t_{2}$ originate from the representation \eqref{ops1}
 for two spins on neighboring tetrahedra (thus $t_{1},t_{2} \sim J' t^{2}$.)
 Finally, $t_3 \sim J' t \Gamma$ reflects  terms of the type $T_{{\bf i}}^{y} ( b_{\bf i}^{\dagger} +
 {\mbox{h.c.}})$,
 i.e. the coupling between the magnetic component of the ground state pseudospin
 and the magnetic excited state (we have set 
$T_{{\bf i}}^{y} \rightarrow \langle T_{{\bf i}}^{y} \rangle =1/2$, as 
\eqref{phamDM} demands.)

%-----------------------
%Fig4
%-----------------------
\begin{figure}[tb]
  \centering
  \includegraphics*[width=0.43\textwidth]{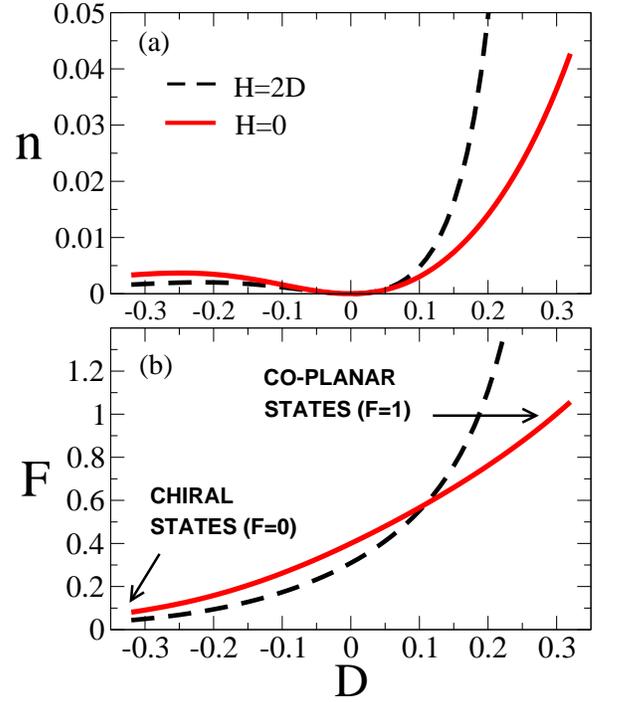}
  \caption{(Color online) (a) The $b$ condensate density $n=|\langle b \rangle|^{2}$
 in zero field (solid red line), and in a finite magnetic field (dashed line).
 (b) The parameter $F$ characterizing the level of coplanarity of the spin order
 at zero field  (solid red line) and in a  finite magnetic field along [001] (dashed line.)
  }
  \label{Fig4}
\end{figure}

From \eqref{teff} it follows that $b$ condenses with a condensate value of
\begin{equation}
\langle b_{\bf i} \rangle =  \langle b \rangle = -\frac{t_3}{\Delta + 4(t_{1}+ t_{2})},
\end{equation}
calculated in the mean-field approximation. 
It is easy to check that $\langle b \rangle < 0$.  The  $b$ bosons are
 hard-core, in order to represent correctly the original spin operators. However
 their hard-core nature can be neglected, and thus the mean-field works well,
 as long as the condensate density
 $n=|\langle b \rangle|^{2}$ is small.  We find this  to be the case for $|D| \lesssim 0.4$;
 beyond this value the boson repulsion would stabilize the rise of $\langle b \rangle$, but
 this regime occurs only for unphysically large values of $D$.
 The density $n$ is plotted in  Fig.~\ref{Fig4}(a)  with a solid red line, and 
 from now on we set  $J'=J$, for definitiveness.
 It is clear that a big difference exists between the two cases
 $D>0$ and $D<0$, partially reflecting the  difference
 in the behavior of  $\Delta$ (Fig.~\ref{Fig2}(a)). For $D>0$ the value
of $\langle b \rangle$ is significantly different from zero, while
 $\langle b \rangle \approx 0$ for $D<0$.

 Let us now investigate how the presence of  $\langle b \rangle \neq 0$
 affects the spin order in the ground state. We combine equations \eqref{chis} 
 and \eqref{ops1}, with $T_{{\bf i}}^{y}=1/2, \  b = \langle b \rangle$,  to obtain
(the site index $i$ is  defined as in Fig.~\ref{Fig1}(c))
\begin{eqnarray}
\label{ops2}
\langle S_{i}^z \rangle & = & (-1)^{i+1}\frac{\Gamma}{2} 
\left ( 1-F \right ) , \  \ i=1,2,3,4,
  \nonumber \\
\langle S_{1}^x \rangle &  =  & \langle S_{2}^x \rangle   = - \langle S_{3}^x \rangle
  = - \langle S_{4}^x \rangle  = - \frac{\Gamma}{2} 
\left ( 1+\frac{F}{2} \right )\!,
   \nonumber \\
\langle S_{1}^y \rangle & = & \langle S_{4}^y \rangle   = - \langle S_{3}^y \rangle
 = - \langle  S_{2}^y \rangle  =  - \frac{\Gamma}{2}
\left ( 1+\frac{F}{2} \right )\!,
\end{eqnarray}
where we have defined
\begin{equation}
F = -\frac{4 \sqrt{2} \ t}{\Gamma} \langle b \rangle = \frac{4 \sqrt{2} \ t}{\Gamma} |\langle b \rangle| \, .
\end{equation}
The function $F$ is plotted in Fig.~\ref{Fig4}(b)(solid red line.)
 For $F=0$ the magnetic order is  of the chiral type (Fig.~\ref{Fig3}(a)), 
 while for $F=1$ it is coplanar, as shown in
Fig.~\ref{Fig3}(b). At the coplanar point the magnetic moment
 is $|\langle {\bf S}_{i} \rangle| = (3\sqrt{2}/4) \Gamma$.
  For an intermediate value of $F$ the order in not coplanar. 
 We find the universal trend   that for $D<0$ the order is almost chiral
 ($F$ is small), and for $D>0$ there is a strong tendency towards coplanar order.

\subsection{Enhancement of coplanar order by magnetic field}
 
 We now proceed to investigate how an external magnetic field can affect 
 the ordering tendencies described above. 
It is known, for  example in dimer systems,  \cite{DimersDM2} that  a field
can induce magnetization perpendicular to it in the presence of DM interactions.
 The exact solution of the problem in our case is rather complex, and below we only give
 a summary of the results for the case of weak fields. We consider  a magnetic field $H$ 
along [001], i.e.  in the $z$ direction in the coordinate system of Fig.~\ref{Fig1}(c). 
 The field is assumed to be weak in the sense that $H \sim D$. 
 The presence of a field leads to changes in the spectrum, and we have found that
 quantitatively the most visible effect is related to the change of the $b$ level energy. \cite{remark}
 In fact  in magnetic field the $b$ state mixes with the state $|P_z \rangle$ (Eq.~\eqref{ex2}),
 and the new eigenstate, $c_{1}|b\rangle+c_{2}|P_z \rangle$, has energy
\begin{equation}
\label{energyh}
E_{b}(H)= -\frac{1}{2} -\frac{3 \sqrt{2}}{4}D - \frac{\sqrt{2}}{4} D
\left[ 1+ \frac{16}{3}\frac{H^{2}}{D^{2}}\right]^{1/2} .
\end{equation}
The  on-site gap in a field, $\Delta_{H} = E_{b}(H) - E_{0}$, is plotted
 in  Fig.~\ref{Fig2}(a) with a dashed line (for the specific choice $H=2D$).
 This behavior in turn enhances the tendency towards coplanar order, as evidenced
 by the dashed lines in Fig.~\ref{Fig4}(a,b). We do not present the exact form 
 of the spin operators in this case, but just mention that they differ slightly
 from the zero-field case \eqref{ops2}. In particular, at the coplanar point,
the magnetic moments are  are not equal on all sites of the tetrahedron,
 as shown in  Fig.~\ref{Fig3}(c), where the blue (upper) arrows are longer
 than the green (lower) ones. 

\section{Conclusions}

 In this work we showed that, in the anisotropic version of the pyrochlore
 lattice and in the presence of DM interactions,  two main types of
 spin order are possible, as summarized in Figure~\ref{Fig3}.
 These are (non-coplanar) chiral and coplanar orderings, depending on the sign
 of $D$. In our model the spin order also generally deviates slightly
from the exact chiral and coplanar configurations, and this deviation itself
 depends on $D$.  It is interesting to note that magnetic order of the kind
 described in this work was also found in the $Sp(N)$ (large $N$) approach to
 the Heisenberg pyrochlore model. \cite{TMS2}
In addition,  collinear order is also possible in that case, while DM interactions do not favor
collinear states. 
 The types of DM-induced order (chiral and coplanar)  we found in the quantum case are also
 consistent with the Monte-Carlo results for the classical (3D)  model. \cite{MCSL}

It is also important to emphasize that our  approach assumes  a specific
 type of lattice-symmetry breaking (as the explicit anisotropy 
 of Fig.~\ref{Fig1}(b) demands.)  In the full (3D) quantum  pyrochlore lattice,
 with Heisenberg interactions only, it is believed that  (spontaneous) lattice-symmetry breaking 
 always takes place in the ground state, \cite{T,BAA,MSG,KEZM} and a certain 
 dimerization pattern sets in. Two-dimensional projections previously studied 
\cite{FMSL,TYM} also exhibit  valence-bond solid ground states.
 In our  quasi-2D version   (Fig.~\ref{Fig1}(b)),
 the dimerization tendency on the tetrahedra is very weak (as it occurs only
 in fourth order of perturbation theory)  and  does not interfere
 with the DM-induced magnetic order formation. At the same time our results 
 are not directly relevant to the 3D pyrochlore lattice.  Also, in this work 
 we have not  provided a dynamical mechanism for the layer decoupling,
 which we have taken as our starting point (the presence of spin-phonon
 interactions with the correct symmetry can certainly accomplish this task.)
 Instead, our goal has been the study  of  time-reversal symmetry breaking
 in certain anisotropic model situations with strong frustration.

Finally, the magnetic order sets in only below the Ising transition
 temperature $T_c \sim J'  \Gamma ^{2}$, as dictated by the pseudospin
 interactions.  Since $\Gamma$  is determined by the DM interaction
 (Fig.~\ref{Fig2}(b)), we expect $T_c$ to be small, and so are the magnetic
 moments  (in all magnetic patterns),
  $|\langle {\bf S}_{i} \rangle| \sim |\Gamma|$.
 So far we are not aware of any convincing evidence that DM-induced
 order takes place in the pyrochlore-related compounds;
 nevertheless the presence of genuine antiferromagnetic order
 is a fundamental property of pyrochlore systems, which
 distinguishes them from other situations with more  ``trivial" manifestations
 of DM interactions (such as weak ferromagnetism.)

\begin{acknowledgments}  
 We are grateful to  M. Elhajal, F. Mila, A. Sandvik,   M. Zhitomirsky,
 and A. H. Castro Neto for valuable discussions related to the topic of this paper.  
\end{acknowledgments}

\appendix
\section{Spin operators on  a tetrahedron}
For completeness we summarize the values of the various matrix
 elements used in the main text. At $D=0$, $|s_1\rangle, |s_2\rangle$ are
 the singlet ground states, and $|t_{\alpha}\rangle,|p_{\alpha}\rangle,|q_{\alpha}\rangle, \ \alpha=x,y,z$
 are $S=1$
 states. The lower left (site) index on the spin operators corresponds
 to the upper sign on the right hand side (and the lower right index
 corresponds to the lower sign, if different.)
\begin{eqnarray}
\label{mel}
&&S_{1,3}^\alpha  =  -\frac{1}{\sqrt{6}}\, t_\alpha^\dagger s_1
\pm \frac{1}{2\sqrt{3}}\, p_\alpha^\dagger s_1
\mp \frac{1}{2}\, q_\alpha^\dagger s_2 + {\rm h.c.} \nonumber  \\
 &  & - \frac{i}{4}\,e^{\alpha\beta\gamma} t_\beta^\dagger t_\gamma
- \frac{i}{2}\,e^{\alpha\beta\gamma} q_\beta^\dagger q_\gamma
\pm  \frac{i}{2\sqrt{2}}\,e^{\alpha\beta\gamma}
(t_\beta^\dagger p_\gamma + p_\beta^\dagger t_\gamma), \nonumber \\
&&S_{2,4}^\alpha  =  \frac{1}{\sqrt{6}}\, t_\alpha^\dagger s_1
\pm \frac{1}{2\sqrt{3}}\, q_\alpha^\dagger s_1
\mp \frac{1}{2}\, p_\alpha^\dagger s_2 + {\rm h.c.} \nonumber \\
&& - \frac{i}{4}\,e^{\alpha\beta\gamma} t_\beta^\dagger t_\gamma
- \frac{i}{2}\,e^{\alpha\beta\gamma} p_\beta^\dagger p_\gamma
\mp \frac{i}{2\sqrt{2}}\,e^{\alpha\beta\gamma}
(t_\beta^\dagger q_\gamma + q_\beta^\dagger t_\gamma). \nonumber \\
\end{eqnarray}
The states are defined as 
\begin{eqnarray}
&&|s_1\rangle   =  \frac{1}{\sqrt{12}} \left[ 
|\uparrow\downarrow\downarrow\uparrow \rangle +
|\downarrow\uparrow \uparrow \downarrow  \rangle +
|\uparrow \uparrow \downarrow\downarrow \rangle  +
|\downarrow\downarrow\uparrow\uparrow \rangle \right. \nonumber \\
&& \left. - 2(|\uparrow\downarrow\uparrow\downarrow \rangle
+ |\downarrow\uparrow\downarrow\uparrow \rangle) \right],
 \nonumber \\
&&|s_2\rangle   =  \frac{1}{2}\left[
|\uparrow\downarrow\downarrow\uparrow \rangle +
|\downarrow\uparrow \uparrow \downarrow \rangle
-|\uparrow \uparrow \downarrow\downarrow \rangle -
|\downarrow\downarrow\uparrow\uparrow \rangle
\right],
\nonumber \\
&&|t_z\rangle  =  \frac{1}{\sqrt{2}}\left[|\uparrow\downarrow\uparrow\downarrow \rangle
-|\downarrow\uparrow\downarrow\uparrow \rangle \right], \nonumber \\
&&|p_z\rangle  =  \frac{1}{2}\left[ |\uparrow\downarrow\downarrow\uparrow \rangle -
|\downarrow\uparrow \uparrow \downarrow \rangle +  |\uparrow \uparrow \downarrow\downarrow
\rangle
-|\downarrow\downarrow\uparrow\uparrow \rangle \right], \nonumber \\
&&|q_z\rangle  =  \frac{1}{2}\left[ -|\uparrow\downarrow\downarrow\uparrow \rangle +
|\downarrow\uparrow \uparrow \downarrow \rangle +  |\uparrow \uparrow \downarrow\downarrow
\rangle
-|\downarrow\downarrow\uparrow\uparrow \rangle \right]. \nonumber \\
\end{eqnarray}
Only the spin zero components of the triplets
 are shown, and in the notation of the type $|\uparrow\downarrow\downarrow\uparrow \rangle$
 the spins follow the order $1,2,3,4$, with the  site labels defined in Fig.~\ref{Fig1}(c).


\begin{thebibliography}{99}
\bibitem{remark} The ground state energy (of the states $| \Phi \rangle$ and $| \Psi \rangle$)
 also changes in a field, but only by amount $\sim D^2H^2$, \cite{KEZM} which  is much
 smaller than the field effect in \eqref{energyh}.
\end{thebibliography}
\end{document}